\documentclass[iop,apj,useAMS,usenatbib]{emulateapj-rtx4}

\usepackage{graphicx}
\usepackage{lscape}
\usepackage{amsmath}
\usepackage{longtable}
\usepackage[usenames,dvipsnames]{color}
\usepackage[textwidth=4cm]{todonotes}
\usepackage{hyperref}

\def\tabcolsep{3pt}

\newcommand{\herschel}{{\it Herschel}}
\newcommand{\hermes}{HerMES}

\newcommand{\chandra}{{\it Chandra}}

\newcommand{\myr}{${\rm M_{\sun}yr^{-1}}$}
\newcommand{\lsun}{${\rm L_{\sun}}$}

\newcommand{\lya}{Ly$\alpha$}
\newcommand{\fesc}{$f_{\rm esc}$(\lya)}

\newcommand{\cii}{[\ion{C}{2}]158\micron}

\def\mnras{MNRAS}
\def\aap{A\&A}

\def\aj{AJ}
\def\araa{ARA\&A}
\def\apjl{ApJ}
\def\apj{ApJ}
\def\apjs{ApJS}
\def\pasp{PASP}
\def\nat{Nature}

\begin{document}

\shorttitle{Far-IR constraints on the Ly$\alpha$ escape fraction}
\shortauthors{J.\,L.\ Wardlow et al.}
\title{Constraining the Lyman Alpha Escape Fraction with Far-Infrared Observations of Lyman Alpha Emitters}
\author{Julie L. Wardlow\altaffilmark{1,2$\dag$},
S.~Malhotra\altaffilmark{3},
Z.~Zheng\altaffilmark{3},
S.~Finkelstein\altaffilmark{4},
J.~Bock\altaffilmark{5,6},
C.~Bridge\altaffilmark{5},
J.~Calanog\altaffilmark{1},
R.~Ciardullo\altaffilmark{7,8},
A.~Conley\altaffilmark{9},
A.~Cooray\altaffilmark{1,5},
D.~Farrah\altaffilmark{10},
E.~Gawiser\altaffilmark{11},
C.~Gronwall\altaffilmark{7,8},
S.~Heinis\altaffilmark{12},
E.~Ibar\altaffilmark{13,14},
R.J.~Ivison\altaffilmark{15},
G.~Marsden\altaffilmark{16},
S.J.~Oliver\altaffilmark{17},
J.~Rhoads\altaffilmark{3},
D.~Riechers\altaffilmark{18},
B.~Schulz\altaffilmark{5,19},
A.J.~Smith\altaffilmark{17},
M.~Viero\altaffilmark{5},
L.~Wang\altaffilmark{20},
M.~Zemcov\altaffilmark{5,6}}
\altaffiltext{$\dag$}{jwardlow@dark-cosmology.dk}
\altaffiltext{1}{Department of Physics \& Astronomy, University of California, Irvine, CA 92697}
\altaffiltext{2}{Dark Cosmology Centre, Niels Bohr Institute, University of Copenhagen, Denmark}
\altaffiltext{3}{School of Earth and Space Exploration, Arizona State University, Tempe, AZ 85287}
\altaffiltext{4}{The University of Texas at Austin, Austin, TX 78712}
\altaffiltext{5}{California Institute of Technology, 1200 E. California Blvd., Pasadena, CA 91125}
\altaffiltext{6}{Jet Propulsion Laboratory, 4800 Oak Grove Drive, Pasadena, CA 91109}
\altaffiltext{7}{Department of Astronomy and Astrophysics, Pennsylvania State University, University Park, PA 16802}
\altaffiltext{8}{Institute for Gravitation and the Cosmos, The Pennsylvania State University, University Park, PA 16802}
\altaffiltext{9}{Center for Astrophysics and Space Astronomy 389-UCB, University of Colorado, Boulder, CO 80309}
\altaffiltext{10}{Department of Physics, Virginia Tech, Blacksburg, VA 24061}
\altaffiltext{11}{Department of Physics and Astronomy, Rutgers, The State University of New Jersey, 136 Frelinghuysen Rd, Piscataway, NJ 08854}
\altaffiltext{12}{Department of Astronomy, University of Maryland, College Park, MD 20742}
\altaffiltext{13}{UK Astronomy Technology Centre, Royal Observatory, Blackford Hill, Edinburgh EH9 3HJ, UK}
\altaffiltext{14}{Universidad Cat\'olica de Chile, Departamento de Astronom\'ia y Astrof\'isica, Vicu\~na Mackenna 4860, Casilla 306, Santiago 22, Chile}
\altaffiltext{15}{Institute for Astronomy, University of Edinburgh, Royal Observatory, Blackford Hill, Edinburgh EH9 3HJ, UK}
\altaffiltext{16}{Department of Physics \& Astronomy, University of British Columbia, 6224 Agricultural Road, Vancouver, BC V6T~1Z1, Canada}
\altaffiltext{17}{Astronomy Centre, Dept. of Physics \& Astronomy, University of Sussex, Brighton BN1 9QH, UK}
\altaffiltext{18}{Department of Astronomy, Space Science Building, Cornell University, Ithaca, NY, 14853-6801}
\altaffiltext{19}{Infrared Processing and Analysis Center, MS 100-22, California Institute of Technology, JPL, Pasadena, CA 91125}
\altaffiltext{20}{Institute for Computational Cosmology, Durham University, South Road, Durham DH1 3LE, UK}
\altaffiltext{$\ddagger$}{{\it Herschel} is an ESA space observatory with science instruments provided by European-led Principal Investigator consortia and with important participation from NASA.}

\label{firstpage}

\begin{abstract}
  We study the far-infrared properties of 498 Lyman Alpha Emitters
  (LAEs) at $z=2.8$, 3.1 and 4.5 in the Extended \chandra\ Deep
  Field-South, using 250, 350 and 500\micron\ data from the
  \herschel\altaffilmark{$\ddagger$} Multi-tiered Extragalactic Survey
  (HerMES) and 870\micron\ data from the LABOCA ECDFS Submillimeter
  Survey (LESS). None of the 126, 280 or 92 LAEs at $z=2.8$, 3.1 and
  4.5, respectively, are individually detected in the far-infrared
  data. We use stacking to probe the average emission to deeper
  flux limits, reaching $1\sigma$ depths of $\sim0.1$ to 0.4\,mJy. The
  LAEs are also undetected at $\ge3\sigma$ in the stacks, although a
  $2.5\sigma$ signal is observed at 870\,\micron\ for the $z=2.8$
  sources. We consider a wide range of far-infrared spectral energy
  distributions (SEDs), including a M82 and an Sd galaxy template, to
  determine upper limits on the far-infrared luminosities and
  far-infrared-derived star-formation rates of the LAEs.  These
  star-formation rates are then combined with those inferred from the
  \lya\ and UV emission to determine lower limits on the LAEs \lya\
  escape fraction (\fesc).
  For the Sd SED template, the inferred LAEs \fesc\ are $\gtrsim30\%$
  ($1\sigma$) at $z=2.8$, 3.1 and 4.5, which are all significantly higher
  than the global \fesc\ at these redshifts. Thus, if the
  LAEs \fesc\ follows the global evolution then they have warmer
  far-infrared SEDs than the Sd galaxy template. The average and
  M82 SEDs produce lower limits on the LAE \fesc\ of $\sim10$--20\%
  ($1\sigma$), all of which are slightly higher than the global evolution of
  \fesc, but consistent with it at the 2--3$\sigma$ level.
\end{abstract}


\section{Introduction}
\label{sec:intro}

The 1216\,\AA\ \lya\ emission line is a tracer of the ionizing photons
radiated by young stars. The spectral line originates from the $\rm{n}
= 2\rightarrow1$ transition of hydrogen and can contain up to
$\sim6\%$ of the bolometric luminosity of a star-forming galaxy
\citep{PartridgePeebles67}. It reliably identifies star-forming
galaxies at redshifts $z>2$, with \lya\ line searches now
well-established as a robust method for selecting samples of
high-redshift galaxies, both using narrowband images
\citep[e.g.][]{CowieHu98, Rhoads00, Rhoads03, Gronwall07, Gawiser07, Ouchi08,
  Finkelstein08, Finkelstein09b, Guaita10} and spectroscopic surveys
\citep[e.g.][]{Steidel99, Deharveng08, Blanc11}. Thousands of
photometrically-selected \lya\ emitters (LAEs) have been identified,
hundreds of which have been spectroscopically confirmed
\citep[e.g.][]{Hu04, Dawson07, Wang09} at $z \approx 0.3$
\citep{Deharveng08, Finkelstein09c, Cowie10} to $z \approx 7$
\citep{Iye06, Ouchi09, Ouchi10, Rhoads12, Shibuya12}.
 
However, the interpretation of \lya\ observations is challenging
because \lya\ photons interact with the neutral hydrogen in the
inter-stellar medium (ISM) and are resonantly scattered. Furthermore,
due to their short wavelength they are also susceptible to absorption
by dust, which further complicates analyses \citep{Neufeld91,
  HansenOh06, Finkelstein09a}. Radiative transfer in the
dusty, multiphase and dynamic interstellar medium (ISM) is complex and
thus observations of the escape fraction of \lya\ photons (\fesc),
defined as the ratio of observed to intrinsic \lya\ emission, are
also useful
for probing the clumpiness and distribution of dust and gas in the
ISM, which is typically spatially unresolved at high redshift.

Various methods have been applied to estimate the intrinsic \lya\
emission, which is required for calculating \fesc. For example, under
the case B recombination theory \citep{BakerMenzel38}, the intrinsic
\lya\ line flux can be estimated using the H$\alpha$ line flux,
corrected for dust extinction; but measuring both the \lya\ and
H$\alpha$ lines is possible only over narrow redshift ranges
(e.g. \citealt{Atek09, Finkelstein11a} for $z\sim 0.3$ LAEs, and
\citealt{Hayes10, Finkelstein11b} for $z\sim 2.3$ LAEs). Other methods
of estimating the intrinsic \lya\ flux rely on \lya\ photons ability
to trace young stars. Thus, intrinsic \lya\ emission is connected to
the intrinsic star-formation rate (SFR), which can be estimated from
the UV continuum (subject to dust extinction), or X-ray emission
\citep{Zheng12} which is extinction free, but relies on an empirically
calibrated relation between X-ray emission and SFR \citep{Nandra02,
  Grimm03, Ranalli03, Persic04,  Symeonidis11}.

Alternatively, the SFR can be estimated from measurements of the
far-infrared continuum emission, which in young galaxies, is emitted
by dust heated by young stars \citep[e.g][]{Kennicutt98, Egami04,
  Choi06, Rieke09, Calzetti10, Murphy11}. The \lya\ and
far-infrared measurements provide complementary views of the non-dusty
and dusty regions of a galaxy, respectively.  In this paper we use
continuum 250, 350, 500 and 870\,\micron\ observations to probe the
dust emission of three samples of LAEs at $z=2.8$ (Zheng et al.\ in
prep.), $3.1$ \citep{Gronwall07, Ciardullo12} and $4.5$
\citep{Finkelstein09b, Zheng13}.  As the LAEs are too faint to be
individually detected in the far-infrared data we use a stacking
analysis to reach deeper flux limits and investigate the average
emission from the sources (c.f.\ \citealt{Davies13} stacking of
$z\sim4.5$ LAEs at 870\micron). Then, by comparing the integrated SFR
derived from the far-infrared luminosity with the integrated SFR
derived from the {\it apparent} \lya\ and rest-frame ultra-violet (UV)
luminosities the \lya\ escape fraction is calculated.

In Section~\ref{sec:data} we present the \lya\ samples and the
far-infrared data used in the analysis. The stacking procedure is
described in Section~\ref{sec:analysis} and the far-infrared SEDs of
LAEs and the \lya\ escape fraction are presented and discussed in
Section~\ref{sec:results}. Our conclusions are presented in
Section~\ref{sec:conc}.  Throughout this paper we use $\Lambda$CDM
cosmology with $\Omega_{\rm M}=0.27$, $\Omega_{\Lambda}=0.73$ and
$H_{0}=71\,{\rm km\,s^{-1}\,Mpc^ {-1}}$.

\section{Data}
\label{sec:data}

\lya\ emission is easily absorbed by dust, and therefore, LAE
selections may preferentially bias against galaxies with bright
far-infrared (dust) emission, although this supposition depends on the
distribution of dust in the ISM. Indeed, we note the large fraction of \lya\
detections amongst SMGs \citep[e.g.][]{Chapman05} and their occasional
association with \lya\ blobs \citep{Ivison98}.
Measurements of dust absorption from
the UV spectral slopes of LAEs also suggest that LAEs will be faint at
far-infrared wavelengths \citep{Finkelstein09a}. Therefore, in this paper we
consider LAEs in the Extended \chandra\ Deep Field South
(ECDFS) survey region, where extensive deep far-infrared data are
available.

\subsection{Sample selection}
\label{sec:sample}

We examine a total of 498 LAEs in the ECDFS in three redshift bins:
$z=2.8$, $z=3.1$ and $z=4.5$.

The $z=2.8$ sample consists of 126 photometrically-selected LAEs
identified in narrowband NB466, NB470, and NB475 (all with FWHM$\sim$
50\AA), with VLT/VIMOS {\it U}-band \citep{Nonino09} and MUSYC {\it B}-band
\citep{Gawiser06} coverage. The selection criteria are $U-B \geq 0.8$,
NB$\geq 5\sigma$, and $B-{\rm NB}\geq 1$ (Zheng et al.\ in prep.).

At $z=3.1$ we examine the 252 and 188 LAEs presented in
\citet{Gronwall07} and \citet{Ciardullo12}, respectively. The samples
are photometrically selected using narrowband imaging with slightly
different filters (for a comparison see \citealt{Ciardullo12}). There
is some overlap between the two catalogs; we remove duplicates using a
matching radius of 1\arcsec, which results in a final sample of 280
unique $z=3.1$ LAEs, of which $\sim 70$ have so-far been
spectroscopically confirmed. 

For the highest redshift sample, at $z=4.5$, we consider LAEs that
were identified in narrowband imaging by \citet{Finkelstein09b}. We
consider the 92 of these LAEs that were confirmed with spectroscopic
followup observations \citep[46 LAEs;][]{Zheng13} or that have not
been spectroscopically targeted (44 LAEs). Our conclusions do not
change if we only consider the 46 spectroscopically confirmed $z=4.5$ LAEs,
although, due to the larger sample size, the stacked flux and SFR
limits are deeper when the photometric LAEs are included.

\begin{figure*}
\centering\includegraphics[width=16cm]{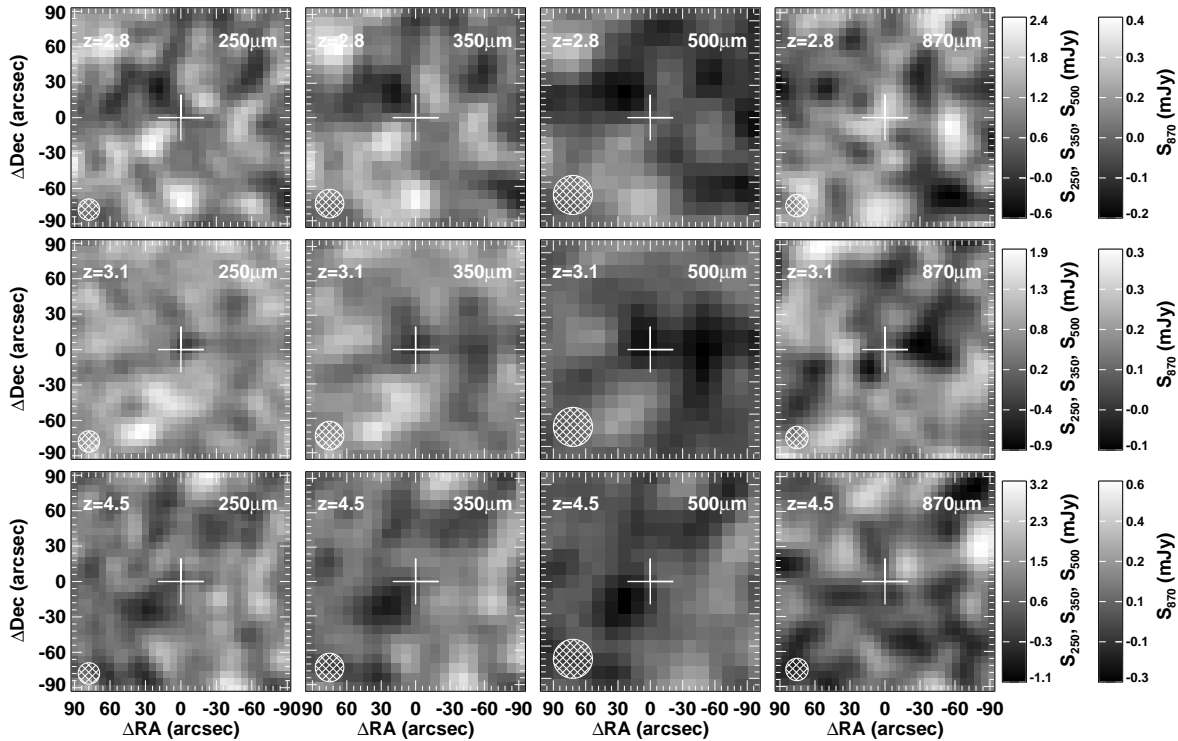}
\vspace{0.1cm}
\caption{$180\times180\arcsec$ cutouts of the $z=2.8$, 3.1 and 4.5
  (top to bottom) stacked LAEs at 250, 350, 500 and 870\,\micron\
  (left to right). The LAEs are positioned at the centers of the
  stacks -- marked with crosses -- and are not detected at $\ge3\sigma$
  in any of the far-infrared data. The shaded circles 
show the size of the beam at each wavelength.}
\label{fig:cut}
\end{figure*}

\begin{figure*}
\centering\includegraphics[width=16cm]{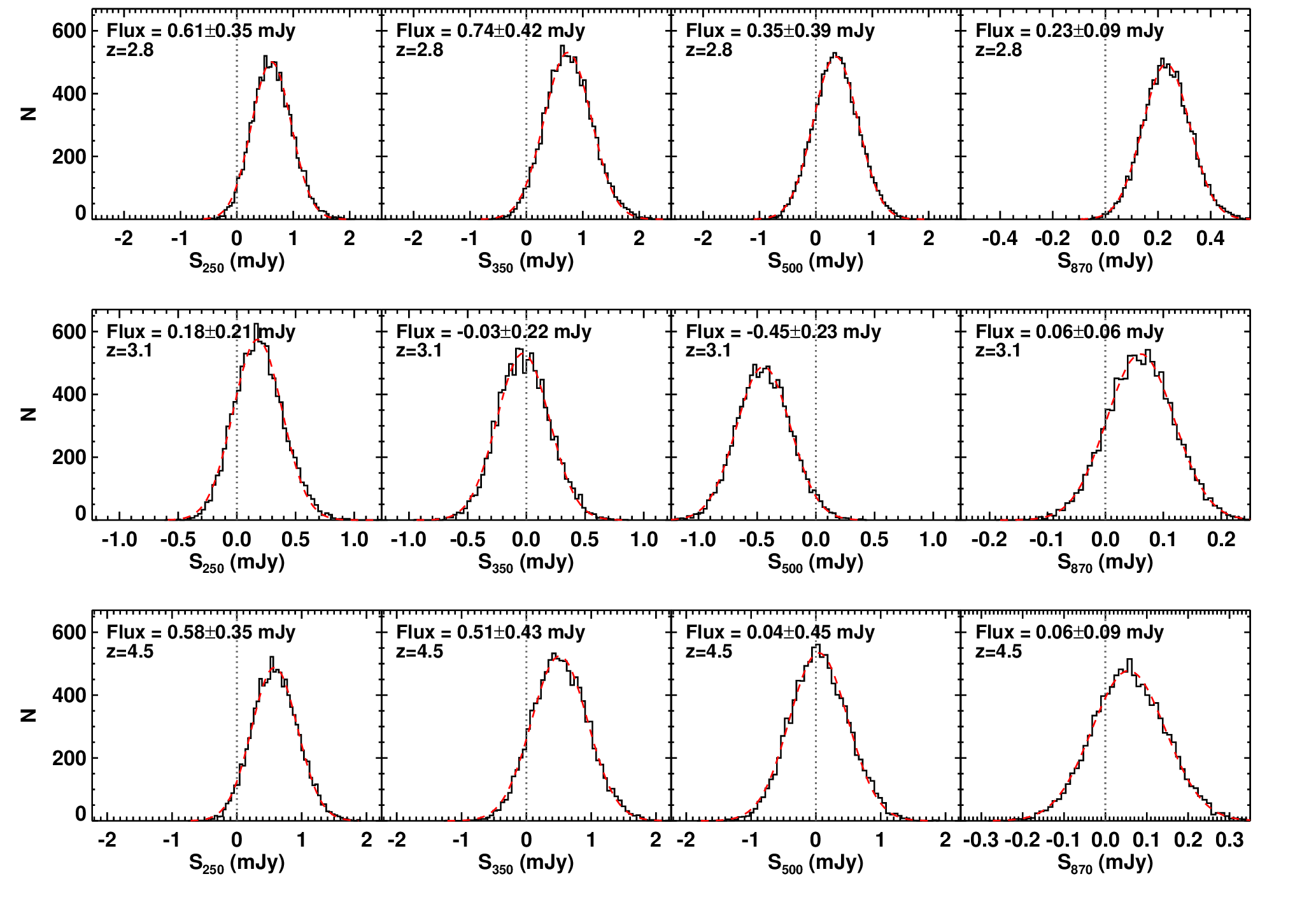}
\caption{Histograms of the 250, 350, 500 and 870\,\micron\ (left to
  right) stacked flux densities measured in 10,000 bootstrap trials of
  $z=2.8$, 3.1 and 4.5 (top to bottom) LAEs. Dashed lines show
  Gaussian fits to the histograms from which we measure the stack flux
  densities and detection limits ($1\sigma$), written at the top of
  each panel. None of the LAEs are significantly ($\ge3\sigma$)
  detected in any of the data, although at $2.6\sigma$ the 870\micron\
  stack of the $z=2.8$ LAEs approaches this threshold. }
\label{fig:histo}
\end{figure*}

\subsection{Far-infrared data}
\label{sec:firdata}

In the far-infrared we consider deep 250-, 350-, 500- and
870-\micron\ continuum imaging. The 870-\micron\ data are from the
LABOCA ECDFS Submillimeter Survey \citep[LESS; ][]{Weiss09} and the
250-, 350- and 500-\micron\ data were taken with SPIRE \citep{Griffin10}
on the {\it Herschel Space Observatory} \citep{Pilbratt10} as part of
the \herschel\ Multi-Tiered Extragalactic
Survey\footnote{\url{http://hermes.sussex.ac.uk}} \citep[HerMES;
][]{Oliver12}.

The LESS 870-\micron\ maps and catalogs are presented in
\citet{Weiss09}. The data cover $30\arcmin\times30\arcmin$, including all
target LAEs, to a roughly uniform depth of
$\sigma\sim1.2$\,mJy\,beam$^{-1}$. These data were taken with the
LABOCA instrument on the 12m APEX telescope resulting in a
19\arcsec\ beam (FWHM). The catalog contains 126 sources down to
$3.7\sigma$, corresponding to $\sim4.4$\,mJy\,beam$^{-1}$.

The HerMES 250, 350 and 500\,\micron\ data in the ECDFS are nested,
with coverage extending over a $204\arcmin\times170\arcmin$ area. All
the target LAEs are located in the central $30\arcmin\times30\arcmin$ of
these data. The central $20\arcmin\times20\arcmin$ region (enclosing
$\sim35\%$ of the LAEs) has the deepest data, reaching down to
$\sigma\sim0.9$, 0.8 and 1.1\,mJy\,beam$^{-1}$ at 250, 350 and
500\micron, respectively (excluding confusion; \citealt{Oliver12}). The remainder of the
central $30\arcmin\times30\arcmin$ reaches $\sigma\sim1.6$, 1.3 and
1.9\,mJy\,beam$^{-1}$ at 250, 350 and 500\micron\ \citep{Oliver12}. For our
analyses all the nested datasets are included and thus the maps and
catalogs have non-uniform coverage. The \herschel\ beam is 18, 25 and
36\arcsec\ (FWHM) at 250, 350 and 500\,\micron, respectively.  Details
of the data reduction and map and catalog production are available in
\citet{Levenson10}, \citet{Viero13b}, \citet{Smith12} and Wang et al.\ (in prep.).

\section{Analysis} 
\label{sec:analysis}

We begin by cross-matching the LAEs with the HerMES and LESS catalogs
to determine whether any are individually detected in the
far-infrared.  The positional uncertainty of the LAEs is typically
$\ll1\arcsec$, which is significantly smaller than that of the
far-infrared catalogs due to the large beamsizes of single dish
submillimeter telescopes. Therefore, the LAE positional uncertainty
can be disregarded when choosing the cross-matching radius and when
stacking the far-infrared data (\S~\ref{sec:stacking}).

For the 19\arcsec\ LESS 870-\micron\ beam the $1\sigma$ positional uncertainty
on the cataloged sources is $\sim 1$--3\arcsec\ \citep{Biggs11,
  Hodge13}, depending on the signal-to-noise ratio (SNR; see
\citealt{Ivison07}). For
cross-matching the LAEs and the LESS sources we choose a liberal
search radius of 9\arcsec -- corresponding to $\sim3\sigma_{\rm pos}$
for the most uncertain positions. Three of the LAEs -- two at $z=2.8$
and one at $z=3.1$ -- are positionally matched to a source in the LESS
catalog, with separations of 6.5 to 8.3\arcsec. None of the $z=4.5$
LAEs are matched in the LESS catalog within the 9\arcsec\ radius.

Assuming that the 498 LAEs and 126 LESS sources are randomly
distributed in the $30\times30\arcmin$ area we expect to find
$\sim1$--2 chance superpositions of LAEs and LESS sources, which is
consistent with all three of LAE---LESS pairs being chance
associations. This interpretation is supported by high-resolution
870\,\micron\ ALMA continuum observations of the 870\,\micron\ LABOCA sources
\citep{Hodge13}, which in two of the cases pinpoints non-LAEs as the
source of the 870\micron\ emission. In the third case, no 870\micron\
sources are detected in the ALMA observations
($\sigma=0.33$\,mJy\,beam$^{-1}$; \citealt{Hodge13}), which may be the
result of blending of several faint far-infrared sources in the LABOCA
beam. In this case the association between the LAE and the LABOCA
source is either a chance superposition, or the LAE does contribute to
the LABOCA source but only a fraction of the detected 870\,\micron\
flux can be from this galaxy. We conclude that none of the
LAEs are robustly individually detected at 870\,\micron.

The 250, 350 and 500\micron\ HerMES catalog is created by blindly
extracting sources at 250\micron, where the beam is smallest
(18\arcsec\ FWHM), using those source positions as priors for the
longer wavelength data and then identifying any additional 350 and
500\micron\ sources in the residual maps. Therefore, the positional
error in the HerMES catalogues is dominated by the 18\arcsec\ beam at
250\micron. The positional error is also
typically $\sim 1$--3\arcsec, depending on the SNR. Therefore, we use
the same liberal 9\arcsec\ search radius when cross-matching the LAEs
with the HerMES catalog. We only consider cataloged sources that are
detected at $\ge3\sigma_{\rm total}$, where $\sigma_{\rm total}$
includes confusion noise, in at least one of the three HerMES bands.
Within the 9\arcsec\ search radius there are two, one and two matches
to the $z=2.8$, 3.1 and 4.5 LAEs, respectively. The five LAE to
HerMES positional matches have separations of 3.4 to
7.9\arcsec. Within the 9\arcsec\ search radius 4--5 chance
superpositions between the $\sim450$ HerMES sources and the 498 LAEs
are expected -- which is consistent with the observed matching rate.
Therefore, it is unlikely that any of the matches between the LAEs
and 250, 350 and 500\micron\ catalog are physical associations
between the LAEs and the far-infrared flux.

\begin{deluxetable*}{ccc|cccc|cccc|cccc}
\tablecaption{Summary of the stacking results 
\label{tab:stats}} 
\startdata 
\hline\hline  
Waveband & Flux density$^a$ & Noise$^b$ & 
\multicolumn{4}{|c}{$1\sigma^{\Upsilon} {\rm L_{IR}}$ ($10^{11}{\rm L_{\sun}}$)} & 
   \multicolumn{4}{|c}{$1\sigma^{\Upsilon}$ SFR (${\rm M_{\sun}\,yr}^{-1}$)}  &  
   \multicolumn{4}{|c}{$1\sigma^{\Lambda}$ \fesc} \\ 
(\micron) & (mJy) & (mJy) & 
Sd$^c$ & M82$^c$ & Mean$^d$ &  All$^e$ &  
Sd$^f$ & M82$^f$ & Mean$^g$ &  All$^h$ &  
Sd$^i$ & M82$^i$ & Mean$^j$ &  All$^k$ \\ 
\hline 
\cutinhead{$z=2.8$; 126 LAEs}
250 &  0.61 &  0.35 & 1.7 & 1.3 & 1.4 & 1 -- 2.1 &  29 &  22 &  24 &  18 -- 36 &  0.07 &  0.09 &  0.08 &  0.06 -- 0.11 \\
350 &  0.74 &  0.42 & 0.94 & 1.5 & 1.2 & 0.9 -- 1.8 &  16 &  26 &  20 &  15 -- 30 &  0.11 &  0.08 &  0.10 &  0.07 -- 0.12 \\
500 &  0.35 &  0.39 & 0.64 & 2 & 1.3 & 0.62 -- 3 &  11 &  34 &  22 &  11 -- 51 &  0.15 &  0.06 &  0.09 &  0.04 -- 0.15 \\
{\bf 870} & {\bf 0.23} & {\bf 0.09} & {\bf0.22} & {\bf1.3} & {\bf0.84} & {\bf0.18 -- 2.5} & {\bf  4} & {\bf 23} & {\bf 15} & {\bf  3 -- 43} & {\bf 0.26} & {\bf 0.09} & {\bf 0.12} & {\bf 0.05 -- 0.28} \\
\cutinhead{$z=3.1$; 280 LAEs}
250 &  0.18 &  0.21 & 1.5 & 0.93 & 1.1 & 0.74 -- 1.7 &  25 &  16 &  19 &  13 -- 30 &  0.06 &  0.08 &  0.07 &  0.05 -- 0.10 \\
350 & -0.03 &  0.22 & 0.68 & 0.94 & 0.78 & 0.59 -- 1 &  12 &  16 &  13 &  10 -- 18 &  0.11 &  0.08 &  0.10 &  0.08 -- 0.12 \\
500 & -0.45 &  0.23 & 0.45 & 1.3 & 0.83 & 0.43 -- 1.8 &   8 &  22 &  14 &   7 -- 30 &  0.15 &  0.06 &  0.09 &  0.05 -- 0.16 \\
{\bf 870} & {\bf 0.06} & {\bf 0.06} & {\bf0.15} & {\bf0.86} & {\bf0.54} & {\bf0.13 -- 1.6} & {\bf  3} & {\bf 15} & {\bf  9} & {\bf  2 -- 27} & {\bf 0.33} & {\bf 0.09} & {\bf 0.13} & {\bf 0.05 -- 0.37} \\
\cutinhead{$z=4.5$; 92 LAEs}
250 &  0.58 &  0.35 & 11 & 4.3 & 6.1 & 3.1 -- 13 & 187 &  75 & 104 &  53 -- 229 &  0.03 &  0.08 &  0.06 &  0.03 -- 0.10 \\
350 &  0.51 &  0.43 & 4.8 & 3.4 & 3.9 & 2.7 -- 6 &  83 &  59 &  67 &  47 -- 103 &  0.07 &  0.09 &  0.08 &  0.06 -- 0.11 \\
500 &  0.04 &  0.45 & 2.2 & 3.5 & 2.7 & 2.1 -- 4 &  38 &  61 &  47 &  36 -- 69 &  0.13 &  0.09 &  0.11 &  0.08 -- 0.13 \\
{\bf 870} & {\bf 0.06} & {\bf 0.09} & {\bf0.32} & {\bf1.3} & {\bf0.82} & {\bf0.3 -- 2.1} & {\bf  6} & {\bf 22} & {\bf 14} & {\bf  5 -- 37} & {\bf 0.31} & {\bf 0.18} & {\bf 0.23} & {\bf 0.13 -- 0.32}
\enddata 
\tablecomments{
$^{\Upsilon}$ Upper limits.
$^{\Lambda}$ Lower limits. \\
$^a$ Observed flux density in the stack; all are insignificant
($<3\sigma$; Section~\ref{sec:stacking}).
$^b$ $1\sigma$ noise (Section~\ref{sec:stacking}).
$^c$ $1\sigma$ upper limit on the 8--1000\,\micron\ far-infrared luminosity
calculated assuming the Sd or M82 SED (Section~\ref{sec:sed}). 
$^d$ Mean $1\sigma$ upper limit on the 8--1000\,\micron\ far-infrared
luminosity of all of the SEDs shown in Fig.~\ref{fig:sed}  (Section~\ref{sec:sed}).
$^e$ $1\sigma$ range of upper limits on the 8--1000\,\micron\ far-infrared luminosity
from all of the SEDs shown in Fig.~\ref{fig:sed}  (Section~\ref{sec:sed}). 
$^f$ $1\sigma$ upper limit on the star-formation rate calculated from the
far-infrared luminosity from the Sd or M82 SED using \citet{Kennicutt98}.
$^g$  Mean $1\sigma$ upper limit on the star-formation rate of all of the
SEDs shown in Fig.~\ref{fig:sed} and calculated from the far-infrared
luminosity using \citet{Kennicutt98}.
$^h$ $1\sigma$ range of upper limits on the star-formation rate from the
range of far-infrared luminosities from all of the SEDs shown in
Fig.~\ref{fig:sed}. 
$^i$ $1\sigma$ lower limit on the \lya\ escape fraction for the SFR
derived from the  Sd or M82 SED (Section~\ref{sec:escape}).
$^j$ $1\sigma$ lower limit on the \lya\ escape fraction for the mean
SFR of all the SEDs  (Section~\ref{sec:escape}).
$^k$ Range of $1\sigma$ lower limits on the \lya\ escape fraction for 
the SFRs of all the SEDs (Section~\ref{sec:escape}).
}
\end{deluxetable*}

\subsection{Stacking}
\label{sec:stacking}

We next stack the far-infrared data at the position of the LAEs, to
explore their average emission at 250, 350, 500 and 870\,\micron. Stacking
probes below the nominal detection limit by reducing the background
noise so that a measure of the average flux density of the
stacked sample can be made \citep[e.g.][]{Peacock00, Serjeant04,
  Marsden09, Pascale09, Ivison09, Bethermin10,
  Viero12, Viero13, Heinis13, Calanog13}.  For a sample of $N$ sources, and in the
absence of clustering, stacking decreases the background noise by
a factor of $\sqrt N$. We include all the LAEs in the stacks because
none are definitively individually detected, although our conclusions
do not change if LAEs with far-IR sources within 9\arcsec\ are
excluded.

We use the public {\sc idl} code\footnote{Available from
  \url{www.ias.u-psud.fr/irgalaxies/downloads.php}} from
\citet{Bethermin10} to perform separate 250, 350, 500 and
870\,\micron\ stacks. In each case a weighted mean stack is performed,
with the weighting equal to the inverse of the error map, which
accounts for the non-uniform depth of the data.  Prior to stacking
the maps are resampled to properly centroid on each of the LAEs, and we
calibrate them so that the median background level is zero. This
increases the flux levels by 1.22, 1.28 and 0.93\,mJy at 250, 350 and
500\,\micron, respectively, and decreases the 870\,\micron\ fluxes by
0.04\,mJy.  Figure~\ref{fig:cut} shows $180\times180\arcsec$ regions
of the stacked maps.

To robustly measure the flux density and associated detection limit in the
stacked maps, we perform bootstrapping with replacement, repeating
each stack 10,000 times with a random sampling of the LAEs each time.
The stacked flux density in each realization is extracted from PSF
fitting to the centers of the stacks.  Figure~\ref{fig:histo} shows
histograms of these flux density values for the each of the 10,000
bootstrap samples. Each histogram is fitted with a gaussian and the
the stacked flux density and $1\sigma$ detection limit are determined from
the center and standard deviation of the gaussian fit, respectively,
as shown on Figure~\ref{fig:histo} and listed in
Table~\ref{tab:stats}.

The average flux densities of the LAEs measured from stacking are
presented in Table~\ref{tab:stats}. None of the LAEs are detected at
$\ge3\sigma$ in any of the far-infrared data. The most significant
flux is from the $z=2.8$ LAEs, which are observed at $2.6\sigma$ in
the 870\micron\ stack. All the other stacks are $<2\sigma$. The
detection limits presented are measured using the method above and are
consistent with the pixel-to-pixel variance in the stacked images
(Figure~\ref{fig:cut}).  For the $z=4.5$ LAEs the 870\,\micron\ limit
is also consistent with the result from \citet{Davies13} who recently
stacked the same LAEs on a source-subtracted LESS map and also found a
non-detection.

The measured flux densities from stacking low-resolution data, such as
those considered here, can be boosted by clustering, due to multiple
sources occupying the far-infrared beam
\citep[e.g.][]{FernandezConde10, Serjeant08, Serjeant10, Bethermin10, Greve10, Kurczynski10,
  Penner11, Viero13}. Accounting for such an effect would decrease the
limits quoted above, and therefore, we disregard this
effect as we have only constrained upper limits on the flux
densities. Furthermore, we note that LAEs are only weakly clustered
\citep[e.g.][]{Ouchi10}, with $r_0\sim2.5$\,Mpc, $r_0\sim 4.6$\,Mpc and
$r_0\sim5.7$\,Mpc, corresponding to
$M_h\sim3\times10^{10}\,M_{\sun}$, $M_h\sim2\times10^{11}\,M_{\sun}$ and
$M_h\sim5\times10^{11}\,M_{\sun}$ at $z=3.1$, $z=4.5$ and $z=5.7$, respectively
\citep{Gawiser07, Kovac07, Ouchi10}, and their surface density is
low (for instance, 0.24\,beam$^{-1}$ for the $z=3.1$ sample and the
  250 and 870\micron\ beam),  meaning that any boosting to the stacked fluxes
from clustering is expected to be small.

\section{Results and discussion}
\label{sec:results}

\subsection{The far-infrared SEDs of LAEs}
\label{sec:sed}

\begin{figure}
\includegraphics[width=8.5cm]{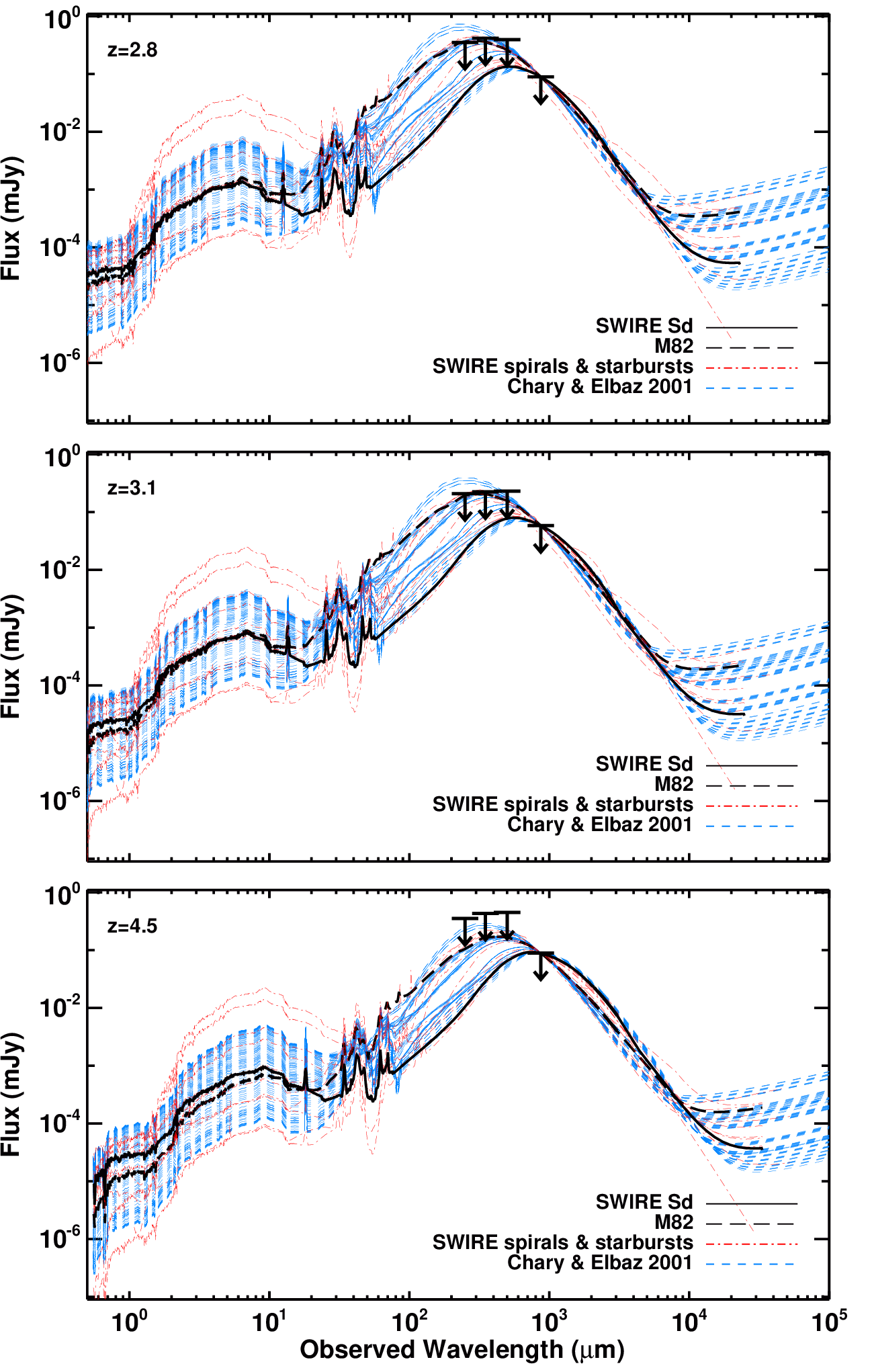}
\caption{Observed 250, 350, 500 and 870\,\micron\ $1\sigma$ stacked
  flux density limits for $z=2.8$, 3.1 and 4.5 (top to bottom) LAEs,
  compared to the SEDs of starburst and late-type galaxies from the
  SWIRE library \citep[][]{Polletta07} and \citet{CharyElbaz01}
  templates. The SEDs are scaled to the $1\sigma$ flux density limit
  at 870~\micron\ and we highlight the Sd and M82 templates for the reasons
  described Section~\ref{sec:sed}. At $z=4.5$ all templates are
  consistent with the observed limits. At $z=2.8$ and $z=3.1$ the 250
  and 350\,\micron\ $1\sigma$ detection limits marginally disfavour
  the warmest SEDs, although none are excluded at the
  $\ge3\sigma$ level, and only a handful at $2\sigma$.  }
\label{fig:sed}
\end{figure}

In Figure~\ref{fig:sed} we show the $1\sigma$ far-infrared flux limits
derived from the stacking in Section~\ref{sec:stacking}, compared to
\citet{CharyElbaz01} SED templates and spirals and starburst galaxies
in the SWIRE compilation \citep{Polletta07}. The templates are all
scaled to the 870\,\micron\ flux density limits, and for the $z=4.5$
sample none violate the 250, 350 or 500\,\micron\ limits. For the
$z=2.8$ and $z=3.1$ galaxies the warmest SEDs violate the $1\sigma$
250 and 350\,\micron\ flux limits, although none are excluded at the
$\ge3\sigma$ level. The $z=2.8$ LAEs
are detected at 1.7, 1.8, and 2.6$\sigma$ significance at 250, 350 and
870\,\micron, respectively, in the stacks. Taking these fluxes and
their associated errors disfavours both the warmest and
the coolest SEDs, including the Sd galaxy template.  We conclude that
LAEs at $z\sim3$ may not be dominated by
the warmest or the coolest dust SEDs, but we cannot constrain the shape of
the LAE's far-infrared SEDs beyond reasonable templates with the
current data. 

It has been suggested that LAEs have dust properties similar to local
Sd galaxies with cooler dust emission than average
\citep{Finkelstein09a}, and thus the SWIRE Sd template is highlighted
in Figure~\ref{fig:sed}. However, recent measurements indicate the
LAEs are typically 1--1.2\,kpc in size \citep{Malhotra12}, which using
the local correlation between star-formation intensity and dust
temperature \citep{Lehnert96}, suggests that LAEs may contain warmer
dust (rest-frame $S_{60}/S_{100}\sim1$) than Sd galaxies. The M82
template in the SWIRE library has $S_{60}/S_{100}\sim1$ and therefore
M82 is also highlighted in Figure~\ref{fig:sed}. The hypothesis that
LAEs contain warmer dust than previously anticipated is consistent
with recent evidence that LAEs have lower metallicities and higher
ionization parameters than LBGs of the same mass \citep[][Song et al.\
in prep.]{Finkelstein11b, McLinden11, Nakajima13, Richardson13}.

\begin{figure*}
\centering
\includegraphics[width=14cm]{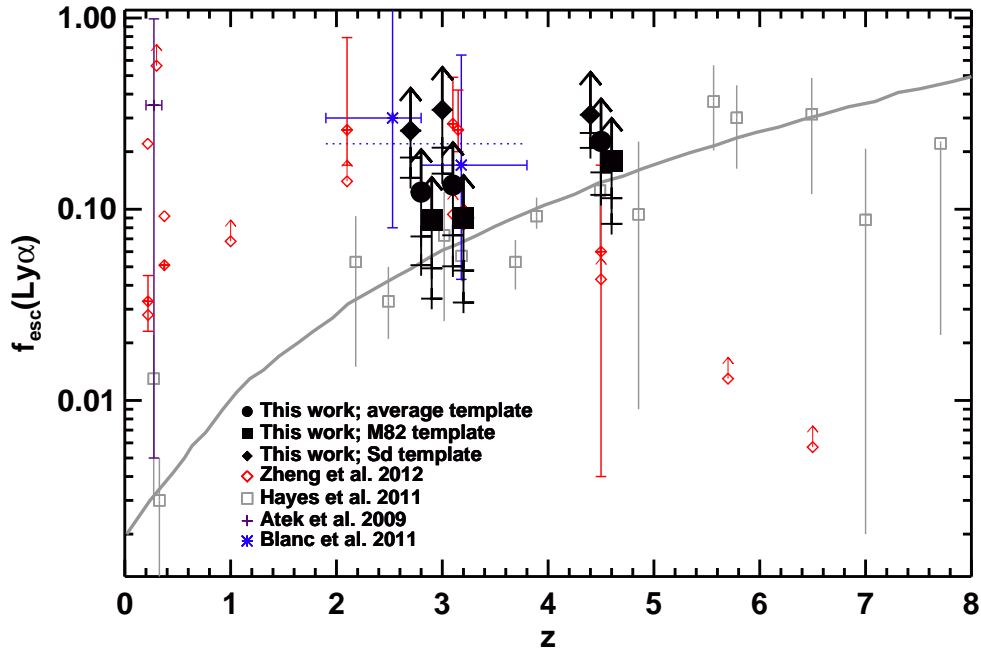}
\caption{The \lya\ escape fraction of LAEs as a function of
  redshift. The large symbols are our $1\sigma$ limits derived for
  three different SED templates from the 870-\micron\ stacking of LAEs
  (with the result for the Sd and M82 templates offset slightly in $z$
  for clarity); the $2\sigma$ and $3\sigma$ limits are shown by the
  lines and tickmarks below each symbol. We compare with X-ray
  stacking results \citep{Zheng12} and spectroscopic and optical
  photometric measurements \citep{Atek09, Blanc11}. The grey points
  and line show the redshift evolution of the {\it global} \lya\
  escape fraction \citep{Hayes11}. At all redshifts the LAE \lya\
  escape fractions that we measure using
  far-infrared emission and the average template or that of M82, are
  consistent with the global evolution at the $\sim2$--$3\sigma$
  level. However, the result using the Sd galaxy templates points to a
  higher \lya\ escape fraction for LAEs than is globally observed.}
\label{fig:escape}
\end{figure*}

Due to the uncertainty in the shape of the typical LAE far-infrared
SED we calculate the $1\sigma$ upper limit on the far-infrared
(8--1000\,\micron) luminosity using both the Sd and M82 templates, as
well as the average luminosity, and the range of luminosities from all
the templates in Figure~\ref{fig:sed}. These values are listed in
Table~\ref{tab:stats} and further illustrate that for a significant
majority of the templates the 870\,\micron\ limit is the most
constraining of the four wavebands examined. In Table~\ref{tab:stats}
we also list the $1\sigma$ limits on the SFRs, calculated from the
far-infrared luminosities of the four wavelengths and four SED types,
using \citet{Kennicutt98}, which assumes a Salpeter initial mass
function (IMF); divide these values by a factor of 1.7 to convert to a
\citet{Chabrier03} IMF. We do not
adjust the SFRs for potential AGN contribution to the far-infrared
emission, because the AGN fraction in LAEs is small
\citep[e.g.][]{Malhotra03, Wang04, Gawiser07}, although the fraction
rises in the lower redshift $z\sim2$ \citep{Nilsson09} and $z\sim0.3$
\citep{Finkelstein09d, Cowie09, Scarlata09} populations.
Furthermore, any adjustment for potential AGN contamination to the
far-infrared fluxes would decrease the stacked flux density limits,
and thus not adjusting these values is the conservative approach.

The Sd template is cooler than the majority of the SEDs, and M82 is
warmer than most of the SEDs. Therefore, using the Sd template and the
870\,\micron\ limit provides lower constraints on the far-infrared
luminosity and the SFR than the M82 template (see also
Figure~\ref{fig:sed}). The mean of all the SEDs lies between the
values provided by these two templates, and thus by considering the
Sd, M82 and mean limits we bracket a wide range of possible LAE
far-infrared SFRs.  At $z=2.8$ we measure $1\sigma$ SFR upper limits
of 4, 23 and 15\,M$_{\sun}$\,yr$^{-1}$ for the Sd, M82 and mean SEDs,
respectively. For the $z=3.1$ LAEs the values are 3, 15 and
9\,M$_{\sun}$\,yr$^{-1}$, and at $z=4.5$ we measure limits of 6, 22
and 14\,M$_{\sun}$\,yr$^{-1}$, respectively.  We note that our results for the
$z=4.5$ LAEs are consistent with \citet{Davies13} who, for the same
sample, calculated SFR$<31$\,M$_{\sun}$\,yr$^{-1}$ ($1\sigma$),
although they only considered the 870\,\micron\ data and assumed a
modified blackbody far-infrared SED with $T_D=35$\,K and $\beta=2.0$.

\subsection{\lya\ Escape Fraction}
\label{sec:escape}

The luminosity of the \lya\ line (${\rm L_{Ly\alpha}}$) can be used to calculate a \lya-derived star-formation rate, SFR$_{\rm Ly\alpha}$, as
\begin{equation} 
{\rm SFR_ {Ly\alpha} (M_{\sun}\,yr^{-1})} = 9.1\times10^{-43}\, {\rm
  L_{Ly\alpha} (erg\,s^{-1}})
\label{eqn:sfrlya}
\end{equation}
for a Salpeter IMF \citep{Kennicutt98, Hu98}. The total star-formation
rate is given by the sum of the unobscured and the dust-obscured
(i.e.\ far-infrared derived) star-formation rates. The \lya\ line is
affected by both dust obscuration and resonant scattering by neutral
hydrogen so we use the apparent (i.e.\ dust-uncorrected) UV luminosity
to trace the unobscured SFR (${\rm SFR_{UV}}$).
Since the intrinsic \lya\ luminosity is also driven by the total
star-formation rate these values can be used to calculate the \lya\
escape fraction (\fesc) as
\begin{equation}
f{\rm _{esc}(Ly\alpha) = SFR_{Ly\alpha} / (SFR_{UV} + SFR_{FIR})},
\label{eqn:fesc}
\end{equation}
where ${\rm SFR_{FIR}}$ is the far-infrared derived (i.e.\ obscured)
SFR for a Salpeter IMF\footnote{For a top-heavy IMF the ratio of ionizing to
non-ionizing photons will be higher, which will preferentially
increase the derived ${\rm SFR_{Ly\alpha}}$
\citep[e.g.][]{FinkelsteinK11} and observed \fesc.}.

For the LAEs in our sample the \lya\ luminosity, and hence SFR$_ {\rm
  Ly\alpha}$, is derived either from flux-calibrated spectroscopy
\citep[e.g.][]{Zheng13} or from the magnitudes of the systems in
narrowband compared to continuum imaging \citep[e.g.][]{Gronwall07,
  Ciardullo12}. For the $z=2.8$, 3.1 and 4.5 LAEs in our analyses the
average SFR$_{\rm Ly\alpha} =2.5$, 1.5 and 7.0\,\myr, respectively,
and the values for ${\rm SFR_{UV}}$ are 6.0, 1.9 and 17\,\myr,
respectively. ${\rm SFR_{UV}}$ is calculated using
  \citet{Kennicutt98} and rest-frame UV luminosities from the observed
 broadband emission minus the effect of the \lya\ line
  \citep[e.g.][]{Ciardullo12,Zheng14}.  Since the UV continuum
  emission is derived from broadband data we also apply a correction for
  attenuation by the intergalactic medium (IGM) of factors of 1.29,
  1.17 and 1.63 to the ${\rm SFR_{UV}}$ at $z=2.8$, 3.1 and 4.5,
  respectively. The IGM correction factors are calculated using
  \citet{Madau95} and the transmission curves of the observed frame
  $B$ ($z=2.8$ sample), $V$ and $B$ ($z=3.1$ sample), and $R$ ($z=4.5$
  sample) filters. We use equation~\ref{eqn:fesc}, the above
  values for SFR$_ {\rm Ly\alpha}$ and ${\rm SFR_{UV}}$, and our
measurement of the upper limit on the far-infrared (i.e.\ dust
obscured) star-formation rates, SFR$_{\rm FIR}$, to calculate lower
limits on the \lya\ escape fraction (Table~\ref{tab:stats}).  
  Note that the corrections for the IGM attenuation of the UV light
  does not affect our conclusions because equation~\ref{eqn:fesc} is
  dominated by our limits on ${\rm SFR_{FIR}}$. As FIR measurements
  get deeper (e.g.\ with ALMA) the IGM attenuation will become more
  important in interpreting studies such as this. 

The 870\,\micron\ data provide the tightest limits of \fesc\ and the
$1\sigma$ limits from these data are shown in Figure~\ref{fig:escape}
and compared with measurements of \fesc\ from LAEs at $z=0$--8 made
using optical spectroscopy and photometry \citep{Atek09, Blanc11} and
X-ray stacking \citep{Zheng12}. We also compare with the global
evolution of \fesc\ measured by \citet{Hayes11}.

The limits on \fesc\ at $z=2.8$, 3.1 and 4.5, calculated using a Sd
galaxy template, are all $>3\sigma$ away from the \citet{Hayes11}
global, optically-derived measurement. This is an indication that
either LAEs have a higher \lya\ \fesc\ than globally observed, or that
they contain warmer dust than typical local Sd galaxies. For the LAEs
in all the redshift bins our far-infrared determinations of \fesc\
using the M82 template are consistent, at the 1--2$\sigma$ level, with
the X-ray results \citep{Zheng12} and the optical determination of the
global \fesc\ from \citet{Hayes11}. For the $z=2.8$ and $z=4.5$ LAEs
the $3\sigma$ limit on the \fesc\ measured using the average
far-infrared SED is at the threshold of being consistent with the
global evolution.

If we consider the $2.6\sigma$ significance detection of the stacked
$z=2.8$ LAEs at 870\,\micron\ (Section~\ref{sec:stacking}) as real,
then the inferred ${\rm SFR_{FIR}}=10\pm4$\,\myr, $58\pm23$\,\myr,
$37\pm15$\,\myr\ (where the errors represent the 870\micron\
photometric uncertainty) for the Sd, M82 and average of the SEDs,
respectively. In this case the inferred \fesc\ are  $0.16\pm0.04$,
$0.04\pm0.01$, and $0.08\pm0.02$, respectively. For the M82 and
average SED these values are consistent with the global \fesc, but for
the Sd galaxy template the inferred LAE \fesc\ is significantly higher
than the global \fesc\ evolution \citep{Hayes11}. Note also, that
the 1--2$\sigma$ significance detections of the $z=2.8$ stacks at 250
and 350\,\micron\ disfavour the Sd SED (see Section~\ref{sec:stacking}).

\subsection{Comparison with previous results}

\citet{Oteo12} cross-matched 56 UV-bright $z=2$--3.5 LAEs with
\herschel-PACS 70, 100 and 160\micron\ catalogs. Of their 56 LAEs four
were detected at 160\,\micron\ ($3\sigma; S_{160}\ge2.0$\,mJy),
indicating ${\rm L_{IR}}\ge10^{12}$\,\lsun\ for \citet{CharyElbaz01}
SEDs -- significantly brighter than the averages of our
samples. However, \citet{Oteo12} did not perform far-infrared analyses
(such as stacking) of their individually-undetected population and
therefore, it is unclear whether the apparent difference between the
samples is due to the UV-bright nature of their LAEs, cosmic variance,
the assumed SEDs, or potentially mis-matching between the PACS source
and the LAEs.

At higher redshift, \citet{Ouchi13} recently failed to detect both
1.2\,mm continuum and the \cii\ emission line from the extended
$z=6.6$ LAE `Himiko' with ALMA.  Using their limit on ${\rm L_{IR}}$
yields \fesc$>0.80$ ($1\sigma$ ) -- significantly higher than expected
from the global evolution \citep[][see also
Figure~\ref{fig:escape}]{Hayes11}, although as Himiko is
  spatially extended the relevant physical effects may be different.
We  also caution that at $z=6.6$ the CMB temperature
($\sim20$\,K) can make it harder to detect a galaxy's dust emission
\citep[e.g.][]{daCunha13}, an effect that \citet{Ouchi13} did not 
  explicitly include in their calculations, and which could increase
the far-infrared luminosity limit and decrease the \fesc\ limit.
However, the high \fesc\ is consistent with the hypothesis that Himiko
has low metallicity and low dust content \citep{Ouchi13}.

Another $z\sim6$ system -- HFLS3 was identified on the basis of its
bright dust emission and does not have a similar metallicity and dust
deficit \citep{Riechers13}. The \lya\ line was not detected in LRIS
spectroscopy but it is in a region of significant skyline
contamination. At $z=4.76$ LESS\,J033229 was also identified on the
basis of bright dust emission, but it is detected in \lya\
\citep{Coppin09} with SFR$_{\rm Ly\alpha}=4$\,\myr, compared to
SFR$_{\rm FIR}\sim1000$\,\myr\ (Swinbank et al.\ 2013) -- indicating
\fesc$\sim0.003$. 
The apparent difference between the \fesc\ measured for
  high-redshift submillimeter galaxies and LAEs is likely a
  selection effect -- submillimeter galaxies are selected on the basis
  of their dust emission and extreme star-formation rates, whereas
  LAEs are identified via the (unobscured) \lya\ emission.

\subsection{Future Prospects}
\label{sec:future}

Having used the deepest available data to probe the far-infrared SEDs
of $z=2.8$, 3.1 and 4.5 LAEs we can place tight limits on the required
depths for future surveys that aim to detect LAEs at far-infrared
wavelengths. Using higher resolution observations, which have lower confusion
limits and can provide deeper data (e.g.\ the 450 and 850\,\micron\
SCUBA-2 Cosmology Legacy
Survey\footnote{\url{www.jach.hawaii.edu/JCMT/surveys/Cosmology.html}}),
or stacking on a larger number of LAEs is required. Alternatively,
interferometric observations targeting individual sources can be used
as their small resolutions can probe below the confusion limit of
single-dish surveys.

At $\sim870$\,\micron\ surveys aiming to detect individual LAEs will
need to probe below our observed $1\sigma$ limits of 0.09, 0.06 and
0.09\,mJy\,beam$^{-1}$ at $z=2.8$, 3.1 and 4.5, respectively.  For
example, continuum mapping with ALMA could reach 0.05\,mJy rms ($\sim$
twice as deep as our stacks) in band 7 (850\,\micron) in just
15\,minutes of integration per source. It
is clear from Figure~\ref{fig:sed} that data shorter than the
far-infrared peak at rest-frame $\sim 60$--100\micron\ are also
required to properly characterize the SEDs and derive accurate
measurements of the far-infrared luminosities, SFRs and hence the
\lya\ escape fraction of LAEs. Ground-based observations are more
challenging at these wavelengths --  for instance, ALMA will take 1.5\,hours per source to
reach 0.2\,mJy\,beam$^{-1}$ in band 9 (500\,\micron) --  meaning that
stacking will still be an attractive prospect to constrain the shape
of the SEDs.

\section{Conclusions}
\label{sec:conc}

We have examined the far-infrared SEDs of 126, 280 and 92 LAEs in the
ECDFS at redshifts 2.8, 3.1 and 4.5, respectively. None of the LAEs
are reliably individually detected in \herschel\ (HerMES) imaging at
250, 350 or 500\,\micron, or in LABOCA (LESS) data at 870\,\micron. 

Therefore, we stacked data at the positions of the LAEs in each
redshift slice to probe deeper into their average far-infrared
emission, reaching $1\sigma=0.09$, 0.06 and 0.09\,mJy at 870\,\micron\
for the $z=2.8$, 3.1 and 4.5 LAEs. The average emission was not
detected at $\ge3\sigma$ in any of the stacks and we find that the
870\,\micron\ flux limits provide the deepest constraints on the LAEs
far-infrared luminosities. We use the 4-band photometric limits to
examine the shape of the LAEs' SEDs, and although the warmest SEDs are
marginally disfavoured the shorter wavelength data are not deep enough
to confidently exclude any.

We calculate upper limits on the far-infrared emission from LAEs at
each redshift using M82, an Sd galaxy and our average galaxy SED
templates. The LAEs have ${\rm L_{IR}\lesssim10^{11}L_{\sun}}$,
although the values vary for the different redshift slices and SED
shapes considered (see Table~\ref{tab:stats}). The luminosity limits
were then used to calculate upper limits on dust-obscured SFRs of LAEs
of a few to a few tens ${\rm M_{\sun}\, yr^{-1}}$ on average. 

Since the far-infrared SFR probes dust-obscured star-formation, and
UV emission probes unobscured star-formation they can be combined
to calculate the total SFR in the LAEs. This total SFR traces the
intrinsic \lya\ luminosity, and we use it to calculate lower limits on
the \lya\ escape fraction for LAEs at $z=2.8$, 3.1 and 4.5. We find
escape fractions of $\gtrsim10\%$ ($1\sigma$) at all the redshifts
considered, although the exact values vary with redshift and the SED
used to calculate the far-infrared luminosity. These limits are
broadly consistent with the global evolution of \fesc\ at the
$\sim1$--3$\sigma$ level, with the exception of the results derived for
the Sd galaxy SED template, where the escape fractions are $>30\%$ in
all cases.

\acknowledgments

We thank Ian Smail and Fabian Walter for helpful discussions and feedback on this
manuscript. 
JLW, AC and DR thank the Aspen Center for Physics for hospitality during the
conception, writing and editing of this paper. This work is supported
in part by the NSF under Grant Numbers PHY-1066293 and AST-1055919.
SM and JR thank the DARK Cosmology Centre and Nordea Fonden in Copenhagen, Denmark, for hospitality during the course of this work.
The Dark Cosmology Centre is funded by the Danish National Research Foundation.
We acknowledge support from the Science and Technology Facilities
Council  [grant number ST/I000976/1].

Based on observations collected at the European Organisation for
Astronomical Research in the Southern Hemisphere, Chile, under
programmes 078.F-9028(A), 079.F-9500(A), 080.A-3023(A), and
081.F-9500(A).
This research has made use of data from the HerMES project
({\url http://hermes.sussex.ac.uk}). HerMES is a Herschel Key Programme
utilizing Guaranteed Time from the SPIRE instrument team, ESAC
scientists and a mission scientist. HerMES is described in
\citet{Oliver12}.  The data presented in this paper will be released
through the \hermes\ Database in Marseille, HeDaM
(\url{http://hedam.oamp.fr/HerMES})
SPIRE has been developed by a consortium of institutes led by Cardiff
Univ. (UK) and including: Univ. Lethbridge (Canada); NAOC (China); CEA,
LAM (France); IFSI, Univ. Padua (Italy); IAC (Spain); Stockholm
Observatory (Sweden); Imperial College London, RAL, UCL-MSSL, UKATC,
Univ. Sussex (UK); and Caltech, JPL, NHSC, Univ. Colorado (USA). This
development has been supported by national funding agencies: CSA
(Canada); NAOC (China); CEA, CNES, CNRS (France); ASI (Italy); MCINN
(Spain); SNSB (Sweden); STFC, UKSA (UK); and NASA (USA).

{\it Facilities:} \facility{APEX (LABOCA)}, \facility{Herschel (SPIRE)}




\end{document}